\documentclass[10pt]{article}

\usepackage{amsmath}
\usepackage{amssymb}

\usepackage{graphicx}

\usepackage{cite}

\usepackage{color} 

\usepackage[labelfont=bf,labelsep=period,justification=raggedright]{caption}

\makeatletter
\renewcommand{\@biblabel}[1]{\quad#1.}
\makeatother

\date{}

\begin{document}

\begin{flushleft}
{\Large
\textbf{A network analysis of countries' export flows: firm grounds for the building blocks of the economy}
}
\\
Guido Caldarelli$^{1,2,3}$, 
Matthieu Cristelli$^{4,\ast}$, 
Andrea Gabrielli$^{2,5}$,
Luciano Pietronero$^{4,5}$,
Antonio Scala$^{1,2}$,
Andrea Tacchella$^{4}$
\\
\bf{1} ISC-CNR - Institute of Complex Systems, Dep. Physics, University of Rome ``Sapienza'', P.le Moro 5, 00185 Rome Italy
\\
\bf{2} LIMS - London Institute for Mathematical Sciences, 22 Audley Street, London UK
\\
\bf{3} IMT - Institutions Market Technology, Piazza S. Ponziano 6, 55100 Lucca Italy
\\
\bf{4} Dep. of Physics, University of Rome ``Sapienza'', P.le Moro 5, 00185 Rome Italy
\\
\bf{5} ISC-CNR - Institute of Complex Systems, Via dei Taurini 19, 00185 Rome Italy
\\
$\ast$ E-mail: matthieu.cristelli@roma1.infn.it
\end{flushleft}

\section*{Abstract}
In this paper we analyze the bipartite network of countries and
products from UN data on country production \cite{HH1,HH2}.  We define
the country-country and product-product projected networks and
introduce a novel method of filtering information based on elements'
similarity.  As a result we find that country clustering reveals
unexpected socio-geographic links among the most competing
countries. On the same footings the products clustering can be
efficiently used for a bottom-up classification of produced goods.
Furthermore we mathematically reformulate the ``reflections method''
introduced by Hidalgo and Hausmann \cite{HH2} as a fixpoint problem;
such formulation highlights some conceptual weaknesses of the
approach.  To overcome such an issue, we introduce an alternative
methodology (based on biased Markov chains) that allows to rank
countries in a conceptually consistent way.  Our analysis uncovers a
strong non-linear interaction between the diversification of a country
and the ubiquity of its products, thus suggesting the possible need of
moving towards more efficient and direct non-linear fixpoint
algorithms to rank countries and products in the global market.


\newpage
\section*{Introduction}

\subsection*{Complex Networks}
Networks emerged in the recent years as the main mathematical tool for
the description of complex systems. In particular, the mathematical
framework of graph theory made possible to extract relevant
information from different biological and social
systems\cite{gcalda,BDGGS07}.  In this paper we use some concepts of
network theory to address the problem of economic
complexity\cite{Jackson09,BMBL09,HM11}.

Such activity is in the track of a long-standing interaction between
economics and physical
sciences\cite{SABGPS02,SB03,SFSVVW09,FPBRMYS05,Amaral09} and it
explains, extends and complements a recent analysis done on the
network of trades between nations\cite{HH1,HH2}.  Hidalgo and Hausmann
(HH) address the problem of competitiveness and robustness of
different countries in the global economy by studying the differences
in the Gross Domestic Product and assuming that the development of a
country is related to different``capabilities''.  While countries
cannot directly trade capabilities, it is the specific combination of
those capabilities that results in different products traded. More
capabilities are supposed to bring higher returns and the accumulation
of new capabilities provides an exponentially growing
advantage. Therefore the origin of the differences in the wealth of
countries can be inferred by the record of trading activities analyzed
as the expressions of the capabilities of the countries.

\subsection*{Revealed Competitive Advantage and the country-product matrix}
We consider here the Standard Trade Classification data for the years
in the interval $1992-2000$.  In the following we shall analyze the
year $2000$, but similar results apply for the other snapshots. For
the year $2000$ the data provides information on $N_{c}=129$ different
countries and $N_{p}=772$ different products.

To make a fair comparison between the trades, it is useful to employ
Balassa's Revealed Comparative Advantage (RCA)\cite{Balassa} i.e. the
ratio between the export share of product $p$ in country $c$ and the
share of product $p$ in the world market
\begin{equation}
RCA_{cp}=\frac{X_{cp}}{{\displaystyle \sum_{p'}}X_{cp'}}/
\frac{{\displaystyle \sum_{c'}}X_{c'p}}{{\displaystyle
    \sum_{c',p'}}X_{c'p'}}\label{eq:RCA}\end{equation} where $X_{cp}$
represents the dollar exports of country $c$ in product $p$.

We consider country $c$ to be a competitive exporter of product $p$ if
its Revealed Comparative Advantage (RCA) is greater than some
threshold value, which we take as 1 as in standard economics
literature; previous studies have verified that small variations
around such threshold do not qualitatively change the results.

The network structure of the country-product competition is given by
the semipositive matrix $M$ defined as

\begin{equation}
M_{cp}=\left\{ \begin{array}{ccc}
1 & if & RCA_{cp}>R^{*}\\
0 & if & RCA_{cp}<R^{*}\end{array}\right.
\label{eq:Mcp}
\end{equation}
where $R^{*}$ is the threshold ($R^{*}$ = 1). 

To such matrix $\hat M$ we can associate a graph whose nodes are
divided into two sets $\{c\}$ of $N_c$ nodes (the countries) and
$\{p\}$ of $N_p$ nodes (the products) where a link between a node $c$
and  a node $p$ exists if and only if $M_{cp}=1$, i.e. a bipartite
graph. The matrix $\hat M$ is strictly related to the adjacency matrix of
the country-product bipartite network.

The fundamental structure of the matrix $\hat M$ is revealed by ordering
the rows of the matrix by the number of exported products and the
columns by the number of exporting countries: doing so, $\hat M$ assumes a
substantially triangular structure.  Such structure reflects the fact
that some countries export a large fraction of all products (highly
diversified countries), and some products appear to be exported by
most countries (ubiquitous products). Moreover, the countries that
export few products tend to export only ubiquitous products, while
highly diversified countries are the only ones to export the products
that only few other countries export.

This triangular structure is therefore revealing us that there is a
systematic relationship between the diversification of countries and
the ubiquity of the products they make. Poorly diversified countries
have a revealed comparative advantage (RCA) almost exclusively in
ubiquitous products, whereas the most diversified countries appear to
be the only ones with RCAs in the less ubiquitous products which in
general are of higher value on the market. It is therefore plausible
that such structure reflects a ranking among the nations.

The fact that the matrix is triangular rather than block-diagonal
suggests that, as countries become more complex, they become more
diversified. Countries add more new products to the export mix 
while keeping, at
the same time, their traditional productions. The structure of $\hat
M$ therefore contradicts most of classical macro-economical models
predicting always a specialization of countries in particular sectors
of production (i.e. countries should aggregate in communities producing
similar goods) that would result in a more or less block-diagonal matrix
$\hat M$.

In the following, we are going to analyze the economical consequences
of the structure of the bipartite country-product graph described by
$\hat M$. In particular, we analyze the community structure induced by
$\hat M$ on the countries and products projected networks.  As a
second step, we reformulate as a linear fixpoint algorithm the HH's
{\em reflection method} to determine the countries and products
respective rankings induced by $\hat M$. In this way we are able to
clarify the critical aspects of this method and its mathematical
weakness. Finally, to assign proper weights to the countries, we
formulate a mathematically well defined biased Markov chain process on
the country-product network; to account for the bipartite structure of
the network, we introduce a two parameter bias in this method. To
select the optimal bias, we compare the results of our algorithm with a
standard economic indicator, the gross domestic product $GDP$. The
optimal values of the parameters suggests a highly non-linear
interaction between the number of different products produced by each
country ({\em diversification}) and the number of different countries
producing each product ({\em ubiquity}) in determining the
competitiveness of countries and products. This fact suggests that, to
better capture the essential features of economical competition of
countries, we need of a more direct and efficient non-linear approach.

\section*{Results}
\subsection*{The network of countries}

In order to obtain an immediate understanding of the economic
relations between countries induced by their products a possible
approach is to define a projection graph obtained from the original
set of bipartite relations represented by the matrix $\hat
M$\cite{bell}.  The idea is to connect the various countries with a
link whose strength is given by the number of products they mutually
produce.  In such a way the information stored in the matrix $\hat M$
is projected into the network of countries as shown in Fig. 1.

The country network can be characterized by the $(N_C \times N_C)$
{\em country-country} matrix $\hat C= \hat M \hat
M^T$.  The non-diagonal elements $C_{cc'}$ correspond to the number of
products that countries $c$ and $c'$ have in common (i.e. are produced
by both countries). They are a measure of their mutual competition,
allowing a quantitative comparison between economic and financial
systems \cite{JL11}; the diagonal elements $C_{cc}$ corresponds to the
number of products produced by country $c$ and are a measure of the
diversification of country $c$.

To quantify the competition among two countries, we can define the
similarity matrix among countries as
\begin{equation}
S^C_{cc'}=2\frac{C_{cc'}}{C_{cc}+C_{c'c'} }\,.
\label{sim1}
\end{equation}
Note that $0\le S^C_{cc'}\le 1$ and that small (large) values indicate
small (large) correlations between the products of the two countries
$c$ and $c'$.  Similar approaches to define a correlation between
vertices or a distance \cite{BCLM03} have often been employed in the
field of complex networks, for example to detect protein correlations
\cite{brun} or to characterize the interdependencies among clinical
traits of the orofacial system \cite{auconi}.

The first problem for large correlation networks is how to visualize
the relevant structure.  The simplest approach to visualize the most
similar vertices is realized by building a Minimal Spanning Tree (MST)
\cite{Mantegna99,MS00}. In this method, starting from an empty graph,
edges $(c,c')$ are added in order of decreasing similarity until all
the nodes are connected; to obtain a tree, edges that would introduce
a loop are discarded. A further problem is to split the graph in
smaller sub-graphs (communities) that share important common feature,
i.e. have strong correlations. {\em Similarity}, like analogous correlation
indicators, can be used to detect the inner structure of a
network; while different methods for community detection vary in their
detailed implementation\cite{GN02,Fortunato10}, they give reasonably
similar qualitative results when the indicators contain the same
information.

The MST method can be thus generalized in order to detect the presence
of communities by adding the extra condition that no edge between two
nodes that have been already connected to some other node is
allowed. In this way we obtain a set of disconnected sub-trees (i.e. a
forest) embedded in the MST. This {\em Minimal Spanning Forest} (MSF)
method naturally splits the network of countries into separate
subsets. This method allows for the visualization of correlations in a
large network and at the same time performs a sort of community detection 
if not precise, certainly very fast.

By visual inspection in Fig.2 we can spot a large subtree composed by
developed countries and some other subtrees in which clear
geographical correlations are present. Notice that each subtree
contains countries with very similar products, i.e. countries that are
competing on the same markets.  In particular, developing countries
seem to be mostly direct competitors of their geographical neighbors.
This is a general feature of economics systems, even if it is not the
most rationale choice\cite{FL99,YZ09}: as an example, both
banks\cite{DIC06} and countries\cite{GL04} trade preferentially with
similar partners, thereby affecting the whole robustness of the
system\cite{PHPUS10,BPPSH10}. This behavior can be reproduced by
simple statistical models based on agents' fitnesses\cite{CCDM02}.

\subsection*{The network of products}
Similarly to countries, we can project the bipartite graph into a
product network by connecting two products if they are produced by the
same one or more countries giving a weight to this link proportional
to the number of countries producing both products. Such network can
be represented by the $(N_P \times N_P)$ {\em product-product} matrix
$\hat P= \hat M^T \hat M$.  The non-diagonal elements $P_{pp'}$ correspond to the
number of countries producing both $p$ and $p'$ have in common, while
the diagonal elements $P_{pp}$ corresponds to the number of countries
producing $p$.

In analogy with Eq.~(\ref{sim1}), the similarity matrix among products
is defined as
\begin{equation}
S^P_{pp'}=2\frac{P_{pp'}}{P_{pp}+P_{p'p'} }\,.
\label{sim2}
\end{equation}
It indicates how much products are correlated on a market: a value
$S^P_{pp'}=1$ indicates that whenever product $p$ is present on the
market of a country, also product $p'$ would be present. This could be
for example the case of two products $p$, $p'$ that are both necessary
for the same and only industrial process.

As in the case of countries, the MSF algorithm can be applied to
visualize correlations and detect communities. In the case of the
product network this analysis brings to an apparently contradictory
results: let's see why.  Products are officially characterized by a
hierarchical topology assigned by UN.  Within this classification
similar issue as ``metalliferous ores and metal scraps'' (groups
27.xx) are in a totally different section with respect to ``non
ferrous metals'' (groups 68.xx).  By applying our new algorithm, 
based on the economical competition network
$\hat M$, one would naively expect that products belonging to the same
UN hierarchy should belong to the same community and {\em vice-versa};
therefore, if we would assign different colors to different UN
hierarchies, one would expect all the nodes belonging to a single
community to be of the same color. In Fig.~3 we show that this is not
the case.  Such a paradox can be understood by analyzing in
closer detail the detected communities with the MSF method.  As an
example, we show in Fig.4 a large community where most of the vertices
belong to the area of ``vehicle part and constituents''.  In this
cluster we can spot the noticeable presence of a vertex belonging to
``food'' hierarchy. This apparent contradiction is solved up by
noticing that such vertex refers to colza seeds, a typical plant
recently used mostly for bio-fuels and not for alimentation: our MSF
method has correctly positioned this "food" product in the "vehicle"
cluster.  Therefore, methods based on community detection could be
considered as a possible rational substitute for current top-down
"human-made" taxonomies\cite{CCDM02}.

\subsection*{Ranking Countries and Products by Reflection Method}

Hidalgo and Haussman (HH) have introduced in \cite{HH1,HH2} the
fundamental idea that the complex set of capabilities of countries (in
general hardly comparable between different countries) can be inferred
from the structure of matrix $\hat M$ (that we can observe).  In this
spirit, ubiquitous products require few capabilities and can be
produced by most countries, while diversified countries possess many
capabilities allowing to produce most products.  Therefore, the most
diversified countries are expected to be amongst the top ones in the
global competition; on the same footing ubiquitous products are likely   
to correspond to low-quality products.

In order to refine such intuitions in a quantitative ranking among
countries and products, the authors of \cite{HH1,HH2} have introduced
two quantities: the $n^{th}$ level \emph{diversification} $d_c^{(n)}$
(called $k_{c,n}$ in \cite{HH1,HH2}) of the country $c$ and the
$n^{th}$ level \emph{ubiquity} $u_p^{(n)}$ (called $k_{p,n}$ in
\cite{HH1,HH2}) of the product $p$.  At the zero$^{th}$ order the
diversification of a country is
simply defined as the number of its products or
\begin{equation}
d_{c}^{(0)}=\sum_{p=1}^{N_{p}}M_{cp}\equiv k_{c}
\end{equation}
where $k_{c}$ is the degree of the node $c$ in the bipartite
country-product network); analogously the zero$^{th}$ order ubiquity 
of a product is defined as the number of
different countries producing it
\begin{equation}
u_{p}^{(0)}=\sum_{c=1}^{N_{c}}M_{cp}\equiv k_{p}
\end{equation}
where $k_{p}$ is the degree of the node $p$ in the bipartite
country-product network.  The diversification $k_c$ is intended to
represent the zero$^{th}$ order measure of the ``quality'' of the
country $c$ with the idea that the more products a country exports the
strongest its position on the marker. The ubiquity $k_p$ is intended
to represent the zero$^{th}$ order measure of the ``dis-value of the
product $p$ in the global competition with the idea that the more
countries produce a product, the least is its value on the market.

In the original approach these two initial quantities are refined in
an iterative way via a so-called ``reflections method'', consisting in
defining the diversification of a country at the $(n+1)^{th}$
iteration as the average ubiquity of its product at the $n^{th}$
iteration and the ubiquity of a country at the $(n+1)^{th}$ iteration
as the average diversification of its producing countries at the
$n^{th}$ iteration:
\begin{equation}
\left\{ 
\begin{array}{c}
d_c^{(n+1)}=\frac{1}{k_c}\sum_{p=1}^{N_p}M_{cp}u_p^{(n)}\\
\\
u_p^{(n+1)}=\frac{1}{k_p}\sum_{c=1}^{N_c}M_{cp}d_c^{(n)}
\end{array}\right.
\label{eq:HHiteration1}
\end{equation}
In vectorial form, this can be cast in the following form
\begin{equation}
\left\{ 
\begin{array}{c}
{\bf d}^{(n)}=\hat J_A{\bf u}^{(n-1)}\\
\\
{\bf u}^{(n)}=\hat J_B {\bf d}^{(n-1)}
\end{array}\right.
\label{eq:HHiteration}
\end{equation}
where ${\bf d}^{(n)}$ is the $N_c-$dimensional vector of components
$d_c^{(n)}$, ${\bf u}^{(n)}$ is the $N_p-$dimensional vector of components
$u_p^{(n)}$, and where we have called $\hat J_A=\hat C \hat M$ and
$\hat J_B= \hat P\hat M^t$ (the upper suffix $t$ stands for
``transpose''), with $\hat C$ and $\hat P$ respectively the $N_c\times
N_c$ and $N_p\times N_p$ square diagonal matrices defined by
$C_{cc'}= k_c^-1 \delta_{cc'}$ and $P_{pp'} = k_p^-1 \delta_{pp'}$.

Such an approach suffers from some flaws. The first one is related to
the fact that the process is defined in a bipartite networks and
therefore even and odd iterations have different meanings. In fact,
let us consider the diversification $d^{(1)}_c$ of the $c^{th}$
country: as prescribed by the algorithm, $d^{(1)}_c$ is the average
ubiquity of the products of the $c^{th}$ country at the $0$-th
iteration.  Therefore countries with most ubiquitous (less valuable)
products would get an highest $1^{st}$ order diversification. On the
other hand, the approximately triangular structure of $\hat M$ tells
us that these countries are the same ones with a small degree and
therefore with a low value of the $0-th$ order diversification ${\bf
  d}^{(0)}$.  As shown to by \cite{HH1,HH2}, this is the case also to
higher orders; therefore the diversifications at even and odd
iterations are substantially an anti-correlated.  Conversely,
successive even iterations are positively correlated so that
$d_c^{(2)}$ looks a refinement of $d_c^{(0)}$, $d_c^{(4)}$ a
refinement of $d_c^{(2)}$ and so on. Same considerations apply to the
iterations for the ubiquity of products.

The major flaw in the HH algorithm is that it is a case of a consensus
dynamics, i.e.  the state of a node at iteration $t$ is just the
average of the state of its neighbors at iteration $t-1$. It is well
known that such iterations have the uniform state (all the nodes
equal) as the natural fixpoint. It is therefore puzzling how such
"equalizing" procedure could lead to any form of ranking. To solve
such a puzzle, let's write the HH algorithm as a simple iterative
linear system and analyze its behavior.

Focusing only on even iterations and on diversifications, 
we can write HH procedure as:
\begin{equation}
{\bf d}^{(2n)}=\hat J_A \hat J_B {\bf d}^{(2n-2)}=(\hat J_A \hat
J_B)^n {\bf d}^{(0)} = \hat H^n {\bf d}^{(0)}\,,
\label{eq:HHfixpoint}
\end{equation} 
where $\hat H=\hat J_A \hat J_B =\hat C\hat M\hat P\hat M^t$ is a
$N_c\times N_c$ squared matrix.  

The matrix $\hat H$ in Eq.\ref{eq:HHfixpoint} is
a Markovian stochastic matrix when it acts {\em from the right} on positive
vectors, in the sense that every element $H_{cc'}\ge 0$ and
\[\sum_{c=1}^{N_c} H_{cc'}=1\,.\]
In particular for the given $\hat M$ adjacency matrix it is also
ergodic. Therefore, its spectrum of eigenvalues is bounded in absolute
value by its unique upper eigenvalue $\lambda_1=1$.
Since $\hat H$ acts on ${\bf d}^{(2n-2)}$ from the left, the
right eigenvector ${\bf e}_1$ corresponding to
the largest eigenvalue $\lambda_1=1$ is simply a uniform vector with
identical components, i.e. in the $n\rightarrow\infty$
limit ${\bf d}^{(2n)}$ converges to the fixpoint ${\bf e}_1$ where all
countries have the same asymptotic diversification. 

It is therefore not a case that HH prescribe to stop their algorithm at a finite number 
of iterations and that they introduce as a recipe to consider as the ranking of 
a country the rescaled version of the $2n^{th}$ level
diversifications \cite{HH2}
\begin{equation}
\tilde d_c^{(2n)} = \frac{d_c^{(2n)}-\overline{d^{(2n)}}}{\sigma_{d}^{(2n)}}\,,
\label{eq:HHrenorm}
\end{equation}
where $\overline{d^{(2n)}}$ is the arithmetic mean of all $d_c^{(2n)}$
and $\sigma_{d}^{(2n)}$ the standard deviation of the same set. 
With these prescription, HH algorithm seems to
converge to an approximately constant value after $\sim 16$
steps. 

This observed behavior can be easily be explained by noticing
that, in contrast with the erroneous statement in \cite{HH2}, finding
the fitness by the reflection method can be reformulated as a
fix-point problem (our Eq. \ref{eq:HHfixpoint}) and solved using the 
spectral properties of a linear system.
In fact,since the ergodic Markovian nature of $\hat H$ we can order 
eigenvalues/eigenvectors such that 
$|\lambda_{N_c}|\le |\lambda_{N_c}|\le ...\le|\lambda_{2}| <\lambda_1=1$. 
Therefore, expanding ${\bf d}^{(0)}$ in terms of the right eigenvectors
$\{{\bf e}_1, {\bf e}_2,..., {\bf e}_{N_c}\}$ of $\hat H$ the initial
condition 
\[  
{\bf d}^{(0)} = a_1 {\bf e}_1 + a_2 {\bf e}_2 + ... +
a_{N_c}{\bf e}_{N_c} ,
\]
we can write the $2n$-th iterate as
\begin{equation}
{\bf d}^{(2n)} = a_1 {\bf e}_1 + a_2 \lambda_2^n {\bf e}_2 +
...+a_{N_c}\lambda_{N_c}^{n}{\bf e}_{N_c} = a_1 {\bf e}_1 + a_2
\lambda_2^n {\bf e}_2 + O \left((\lambda_3/\lambda_2)^n \right)\,.
\end{equation}
Therefore, at sufficiently large $n$ the ordering of the countries is
completely determined by the components of ${\bf e}_2$; notice that
such an asymptotic ordering is independent from the initial condition
${\bf d}^{(0)}$ and therefore should be considered as the appropriate
fixpoint renormalized fitness ${\bf d}^*$ for all countries.

What happens to the HH scheme? At sufficiently large $n$, $\left<{\bf
  d}^{(2n)}\right> \approx a {\bf e}_1$ and $\sigma_{{\bf d}^{(2n)}}
\propto a_2\lambda_2^n {\bf e}_2+
0\left((\lambda_3/\lambda_2)^n\right) $; therefore ${\bf d}^{(2n)}$
becomes proportional to ${\bf e}_2$ (Eq.~\ref{eq:HHrenorm}).  The
number of iterations $it$ needed to converge is given by the ratio
between $\lambda_2$ and $\lambda_3$ ($(\lambda_3 / \lambda_2)^{it} \ll
1$; therefore the $it \sim 16$ iterations prescribed by HH are not a
general prescription but depend on the structure of the network
analyzed.

Notice also that when the numerical reflection method is used, the
renormalized fitness represents a deviation $O(\lambda_2^n)$ from a
constant and can be detected only if it is bigger than the numerical
error; therefore only "not too big" $it$ can be employed.  On the other
hand, the spectral characterization we propose does not suffer from
such a pitfall even when.
Similar considerations can be developed for the even iterations of the
reflection method for the products.

\subsection*{Biased Markov chain approach and non-linear interactions}  

Having assessed the flaws of HH's method, we investigate the
possibility of defining alternative linear algorithms able to
implement similar economical intuitions about the ranking of the
countries while keeping a more robust mathematical foundation.  In
formulating such a new scheme we will keep the approximation of
linearity for the iterations even though we shall find in the results
hints of the non-linear nature of the problem.
 
Our approach is inspired to the well-known PageRank
algorithm \cite{brin}. PageRank (named after the WWW, where vertices
are the pages) is one of the most famous of
Bonacich centrality measures\cite{Bona}.  In the original PageRank
method the ranking of a vertex is proportional to the time spent on it
by an unbiased random walker (in different
contexts\cite{FPBRMYS05} analogous measures assess the stability of a
firm in a business firm network).  

We define the weights of vertices to be proportional to the time 
that an {\em appropriately biased} random walker on the network spends on
them in the large time limit \cite{ZGC10}.  As shown below,
such weights, being the generalization of $k_c$ and $k_p$, give a
measure respectively of competitiveness of countries and
``dis-quality'' (or lack of competitiveness) of products. As the nodes
of our bipartite network are entities that are logically and conceptually separated
(countries and products), we assign to the random walker a different
bias when jumping from countries to products respect to jumping from
products to countries.

Let us call $w_c^{(n)}$ weight of country $c$ at the $n^{th}$
iteration and $w_p^{(n)}$ fitness of product $p$ at the $n^{th}$
iteration. We define the following Markov
process on the country-product bipartite network
\begin{equation}
\left\{
\begin{array}{l}
w_c^{(n+1)}(\alpha,\beta)=\sum_{p=1}^{N_p}G_{cp}(\beta)w_p^{(n)}(\alpha,\beta)\\
\\
w_p^{(n+1)}(\alpha,\beta)=\sum_{c=1}^{N_c}G_{pc}(\alpha)w_c^{(n)}(\alpha,\beta)
\end{array}
\right.
\label{markov-chain}
\end{equation}
where the Markov transition matrix $\hat G$ is given by
\begin{equation}
\left\{
\begin{array}{l}
G_{cp}(\beta)=\frac{M_{cp}k_c^{-\beta}}{\sum_{c'=1}^{N_c}M_{c'p}k_{c'}^{-\beta}}\\
\\
G_{pc}(\alpha)=\frac{M_{cp}k_p^{-\alpha}}{\sum_{p'=1}^{N_p}M_{cp'}k_{p'}^{-\alpha}}
\end{array}
\right.
\label{markov-chain}
\end{equation}
Here $G_{cp}$ gives the probability to jump from product $p$ to
country $c$ in a single step, and $G_{pc}$ the probability to jump
from country $c$ to product $p$ also in a single step. Note that
Eqs.(\ref{markov-chain}) define a $(N_c+N_p)-$dimensional connected
Markov chain of period two. Therefore, random walkers
initially starting from countries, will be found
on products at odd steps and on countries at even ones; the reverse happens 
for random walkers starting from products.  
By considering separately the random walkers starting from countries and from products,
we can reduce this Markov chain
to two ergodic Markov chains of respective dimension $N_c$ and $N_p$. 
In particular, if the walker starts from a country, using a vectorial 
formalism, we can write for the weights of countries
\begin{equation}
{\bf w}_{c}^{(n+1)}(\alpha,\beta)=\hat T(\alpha,\beta)
{\bf w}_{c}^{(n)}(\alpha,\beta)\;\;\;
\label{country-rank}
\end{equation}
where the $N_c\times N_c$ ergodic stochastic matrix $\hat T$ is defined by
\begin{equation}
T_{cc'}(\alpha,\beta)=\sum_{p=1}^{N_p}G_{cp}(\beta)G_{pc'}(\alpha)\,.
\label{T}
\end{equation}
At the same time for products we can write
\begin{equation}
{\bf w}_{p}^{(n+1)}(\alpha,\beta)=\hat S(\alpha,\beta)
{\bf w}_{p}^{(n)}(\alpha,\beta)\,,
\label{product-rank}
\end{equation}
where the $N_p\times N_p$ ergodic stochastic matrix $\hat S$ is given by
\begin{equation}
S_{pp'}(\alpha,\beta)=\sum_{c=1}^{N_c}G_{pc}(\alpha)G_{cp'}(\beta)\,.
\label{S}
\end{equation}
Given the structure of $\hat T$ and $\hat S$, it is simple to show that
the two matrices share the same eigenvalue spectrum which is upper
bounded in modulus by the unique eigenvalue $\mu_1=1$. For both matrices, the
eigenvectors corresponding to $\mu_1$ are the stationary and
asymptotic weights $\{w_c^*(\alpha,\beta)\}$ and
$\{w_p^*(\alpha,\beta)\}$ of the Markov chains.  In order to find
analytically such asymptotic values, we apply the detailed balance
condition:
\begin{equation}
G_{pc}w_c^*=G_{cp}w_p^* \;\;\;\;\forall (c,p)
\label{detailed}
\end{equation}
which gives
\begin{equation}
\left\{
\begin{array}{l}
w_c^*= A \left(\sum_{p=1}^{N_p}M_{cp}k_p^{-\alpha}\right)k_c^{-\beta}\\
\\
w_p^*= B \left(\sum_{c=1}^{N_c}M_{cp}k_c^{-\beta}\right)k_p^{-\alpha}
\end{array}
\right.
\label{biased-rank}
\end{equation}
where $A$ and $B$ are normalization constants.  Note that for
$\alpha=\beta=0$ Eq.~(\ref{markov-chain}) gives the completely
unbiased random walk for which $\hat T=\hat H^t$ where $\hat H$ is
given in Eq.~(\ref{eq:HHfixpoint}). 
Therefore, in this case Eqs.~(\ref{biased-rank}) become
\begin{equation}
\left\{
\begin{array}{l}
w_c^*(0,0)\sim k_c \\
\\
w_p^*(0,0)\sim k_p\,,
\end{array}
\right.
\label{rank-unbiased}
\end{equation}
as for the case of unbiased random
walks on a simple connected network  the asymptotic weight of a node, 
is proportional to its connectivity. Thus, in the case of $\alpha=\beta=0$ we
recover the zero$^{th}$ order iteration of the HH's reflection method. Note that,
in the same spirit of HH, $w_c^*(0,0)$ gives a rough measure of the
competitiveness of country $c$ while $w_p^*$ gives an approximate
measure of the dis-quality in the market of product $p$. By
continuity, we associate the same meaning of competitiveness/disquality 
to the stationary states  $w_c^*$/$w_p^*$ at different values of
$\alpha$ and $\beta$.

To understand the behavior of our ranking respect to the bias, we have
analyzed the mean correlation (square of the Pearson coefficient) for the year 1998 
(other years give analogous results) between the logarithm
of the GDP\footnote{We are aware that GDP is not an absolute measure
  of wealth \cite{ADGMO10} as it does not account directly for
  relevant quantities like the wealth due to natural resources
  \cite{Dasgupta09}). Nevertheless, we expected that GDP monotonically
  increases with the wealth.  What network analysis shows is that the
  number of products is correlated with both quantities.  We envisage
  such kind of analysis in order to define suitable policies for
  underdeveloped countries \cite{Dasgupta10}.}  of each country and
its weight (Eqs.~(\ref{biased-rank}) for different values of $\alpha$
and $\beta$ (see Fig.~\ref{Fig7}).

It is interesting to note that the region of large correlations (region inside
the contour plot in the Fig.~\ref{Fig7}) is found in the positive quadrant 
for about $0.2<\alpha<1.8$ and $0.5<\beta<1$; 
in particular the maximal value is approximately at
$\alpha\simeq 1.1$ and $\beta\simeq 0.8$.  
These results can be
connected with the approximately ``triangular'' shape of the matrix
$\hat M$.  In fact, let us rewrite Eqs.~(\ref{biased-rank})
(apart from the normalization constant) as:
\[
\left\{
\begin{array}{l}
w_c^*\sim k_c^{1-\beta}\left<k_p^{-\alpha}\right>_c\\
\\
w_p^*\sim k_p^{1-\alpha}\left<k_c^{-\beta}\right>_p
\end{array}
\right.  \,,\] 
where $\left<k_p^{-\alpha}\right>_c$ is the
arithmetic average of $k_p^{-\alpha}$ of the products exported by
country $c$ and $\left<k_c^{-\beta}\right>_p$ is the arithmetic average
of $k_c^{-\beta}$ for countries exporting product $p$.  Since $\beta$
is substantially positive and slightly smaller of $1$ and $\alpha$ is
definitely positive with optimal values around $1$, the competitive
countries will be characterized by a good balance between a high value
of $k_c$ and a small typical value of $k_p$ of its products.
Nevertheless, since the optimal values of $\alpha$ are distributed up
to the region of values much larger than 1 (i.e. $1-\beta$ is
significantly smaller than $1$), we see that the major role for the
asymptotic weight of a country is played by the presence in its
portfolio of un-ubiquitous products which alone give the dominant
contribution to $w_c^*$.  A similar reasoning leads to the conclusion
that the dis-value (or ugliness) of a product is basically determined by the
presence in the set of its producers of poorly diversified countries that are basically
exporting only products characterized by a low level of complexity.  

Our new approach based on biased Markov chain theory permits thus to 
implement the interesting ideas developed by HH in \cite{HH2}, 
on a more solid mathematical basis using the framework of linear iterated 
transformations and avoiding the indicated flaws of HH's ``reflection
method''. Interestingly, our results reveal a strongly
non-linear entanglement between the two basic information one can
extract from the matrix $\hat M$: diversification of countries and
ubiquity of products.  In particular, this non-linear relation makes
explicit an almost extremal influence of ubiquity of products on the
competitiveness of a country in the global market: 
having ``good'' or complex products 
in the portfolio is more important than to have many
products of poor value. Furthermore, the information that a product has 
among its producers some poorly diversified countries is nearly
sufficient to say that it is a non-complex (dis-valuable) product in the market. 
This strongly non-linear entanglement between diversifications of countries
and ubiquities of products is an indication of the necessity to go beyond 
the linear approach in order to introduce more sound and direct description 
of the competition of countries and products
possibly based on a suitable {\em ab initio} non-linear approach
characterized by a smaller number of {\em ad hoc} assumptions
\cite{dellidelli}.

\section*{Discussion}

In this paper we applied methods of graph theory to the analysis of
the economic productions of countries. The information is available in
the form of an $N_c \times N_p$ rectangular matrix $\hat M$ giving the
different production of the possible $N_p$ goods for each of the $N_c$
countries. The matrix $\hat M$ corresponds to a bipartite graph, the
country-product network, that can be projected into the
country-country network $C$ and the product-product network $P$.  By
using complex-networks analysis, we can attain an effective filtering
of the information contained in $C$ and $P$. We introduce a new
filtering algorithm that identifies communities of countries
with similar production.  As an unexpected result, this analysis shows
that neighboring countries tend to compete over the same markets
instead of diversifying.  We also show that a classification of goods
based on such filtering provides an alternative product
taxonomy determined by the countries' activity.  We then study the
ranking of the countries induced by the country-product bipartite network.  
We first show that HH's reflection method's ranking is the
fix-point of a linear process; in this way we can avoid  some logical and
numerical pitfalls and clarify some of its weak theoretical points.
Finally, in analogy with the Google PageRank algorithm, we
define a biased, two parameters Markov chain algorithm to assign ranking
weights to countries and products by taking into account 
the structure of the adjacency matrix of the country-product bipartite network.
By correlating the fix-point ranking 
(i.e. competitiveness of countries and products)
with the GDP of each country, we find that the optimal bias parameters of
the algorithm indicate a strongly non-linear interaction 
between the diversification of the countries and 
the ubiquity of the products.

\section*{Materials and Methods}
\subsection*{Graphs}

A graph is a couple $G=(V,E)$ where $V=\left\{ v_{i}|i=1\ldots
n_{A}\right\} $ is the set of vertices, and $E\subseteq V\times V$ is
the set of edges. A graph $G$ can be represented via its adjacency
matrix $A$

\begin{equation}
A_{ij}=\left\{ \begin{array}{cc}
1 & \mbox{if an edge exists between } v_{i} \mbox{ and } v_{j}\\
0 & \mbox{otherwise}\,.
\end{array}\right.
\end{equation}

The degree $k_i$ of the node $v_i$ is the number $\sum_j A_{ij}$ of
its neighbors.

An unbiased random walk on a graph $G$ is characterized by a
probability $p_{ij}=1/k_i$ of jumping from a vertex $v_i$ to one of
its $k_i$ neighbors and is described by the jump matrix
\begin{equation}
J_G = K^{-1} A\,,
\end{equation}
where $K$ is the diagonal matrix $K_{ij}=k_i \delta_{ij}$
corresponding to the nodes degrees.

\subsection*{Bipartite Graphs}
A bipartite graph is a triple $G=(A,B,E)$ where $A=\left\{
a_{i}|i=1\ldots n_{A}\right\} $ and $B=\left\{ b_{j}|j=1\ldots
n_{B}\right\} $ are two disjoint sets of vertices, and $E\subseteq
A\times B$ is the set of edges, i.e.  edges exist only between
vertices of the two different sets $A$ and $B$.

The bipartite graph $G$ can be described by the matrix $\hat M$
defined as
\begin{equation}
M_{ij}=\left\{ \begin{array}{cc}
1 & \mbox{if an edge exists between } a_{i} \mbox{ and } b_{j}\\
0 & \mbox{otherwise}\,.\end{array}\right.
\end{equation}
In terms of $\hat M$, it is possible to define the adjacency matrix
${\cal A}$ of $G$ as
\begin{equation}
{\cal A}=\left[\begin{array}{cc}
0 & M\\
M^{T} & 0\,.\end{array}\right]
\end{equation}.
It is also useful to define the co-occurrence matrices $P^{A}=MM^{T}$
$ $and $P^{B}=M^{T}M$ that respectively count the number of common
neighbors between two vertices of $A$ or of $B$. $P^{A}$ is the
weighted adjacency matrix of the co-occurrence graph $C^{A}$ with
vertices on $A$ and where each non-zero element of $P^{A}$ corresponds
to an edge among vertices $a_{i}$ and $a_{j}$ with weight
$P_{ij}^{A}$.  The same is valid for the co-occurrence matrix $P^{B}$
and the co-occurrence graph $C^{B}$.

Many projection schemes for a bipartite graph $G$ start from
constructing the graphs $C^{A}$ or $C^{B}$ and eliminating the edges
whose weights are less than a given threshold or whose statistical
significance is low.

\subsection*{Matrix from RCA}
To make a fair comparison between
the exports, it is useful to employ Balassa's Revealed Comparative Advantage
(RCA)\cite{Balassa} i.e. the ratio between the export share of
product $p$ in country $c$ and the share of product $p$ in the
world market

\begin{equation}
RCA_{cp}=\frac{X_{cp}}{{\displaystyle \sum_{p'}}X_{cp'}}/
\frac{{\displaystyle \sum_{c'}}X_{c'p}}{{\displaystyle \sum_{c',p'}}X_{c'p'}}\label{eq:RCA}\end{equation}
where $X_{cp}$ represents the dollar exports of country $c$ in product
$p$. 

The network structure is given by the country-product
adjacency matrix $\hat M$ defined as

\begin{equation}
M_{cp}=\left\{ \begin{array}{ccc}
1 & if & RCA_{cp}>R^{*}\\
0 & if & RCA_{cp}<R^{*}\end{array}\right.
\label{eq:Mcp}
\end{equation}
where $R^{*}$ is the threshold. 
A positive entry, $M_{cp}=1$ tells us that
country $c$ is a competitive exporter of the product $p$.

\section*{Acknowledgments}
We thank EU FET Open project FOC nr.255987 and CNR-PNR National Project "Crisis-Lab" for support. 
\section*{Author Contributions}
All the Authors contributed equally to the work
\bibliography{template}
\ \newpage

\begin{figure}
\includegraphics[scale=.6]{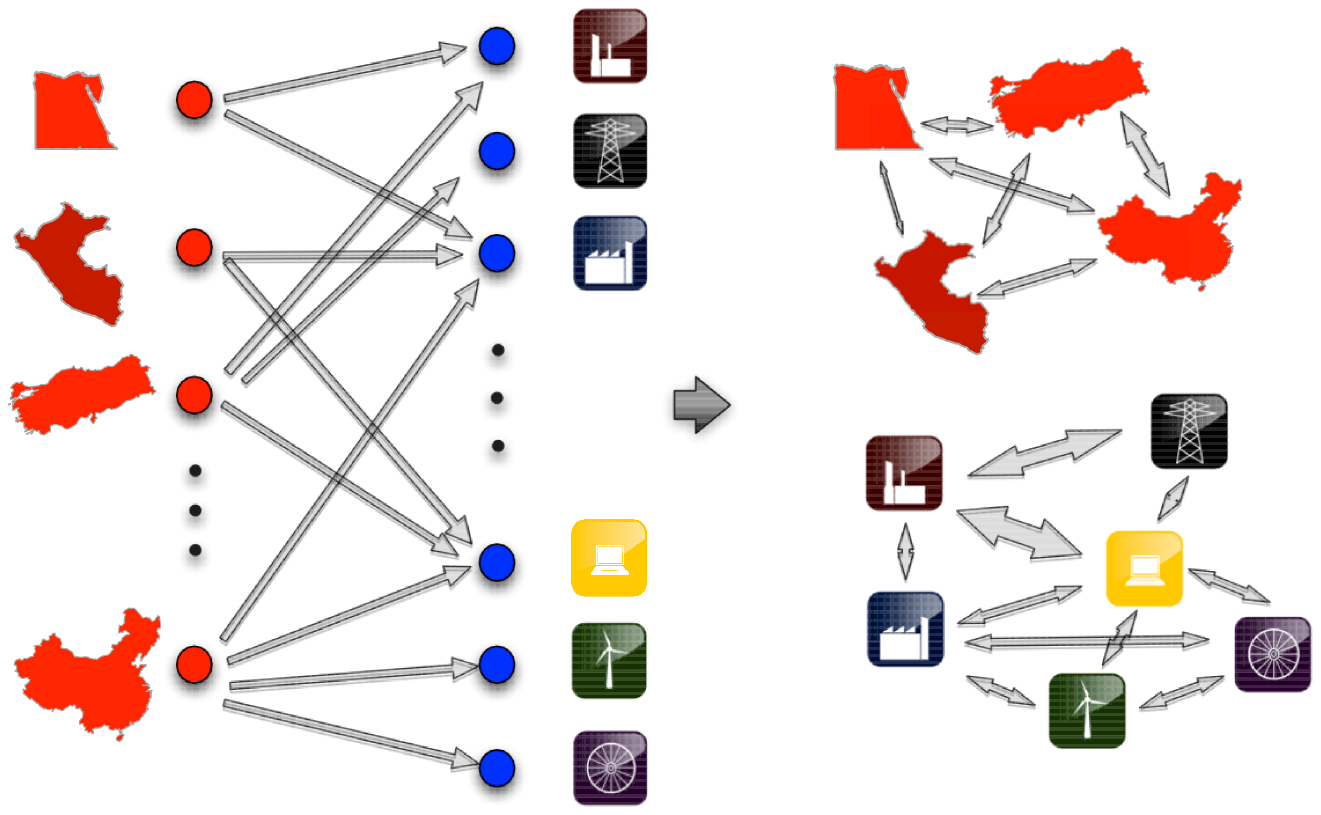}
\caption{\textbf{The network of countries and products and the two possible projections.}}
\label{Fig1}
\end{figure}

\begin{figure}
\includegraphics[scale=.5]{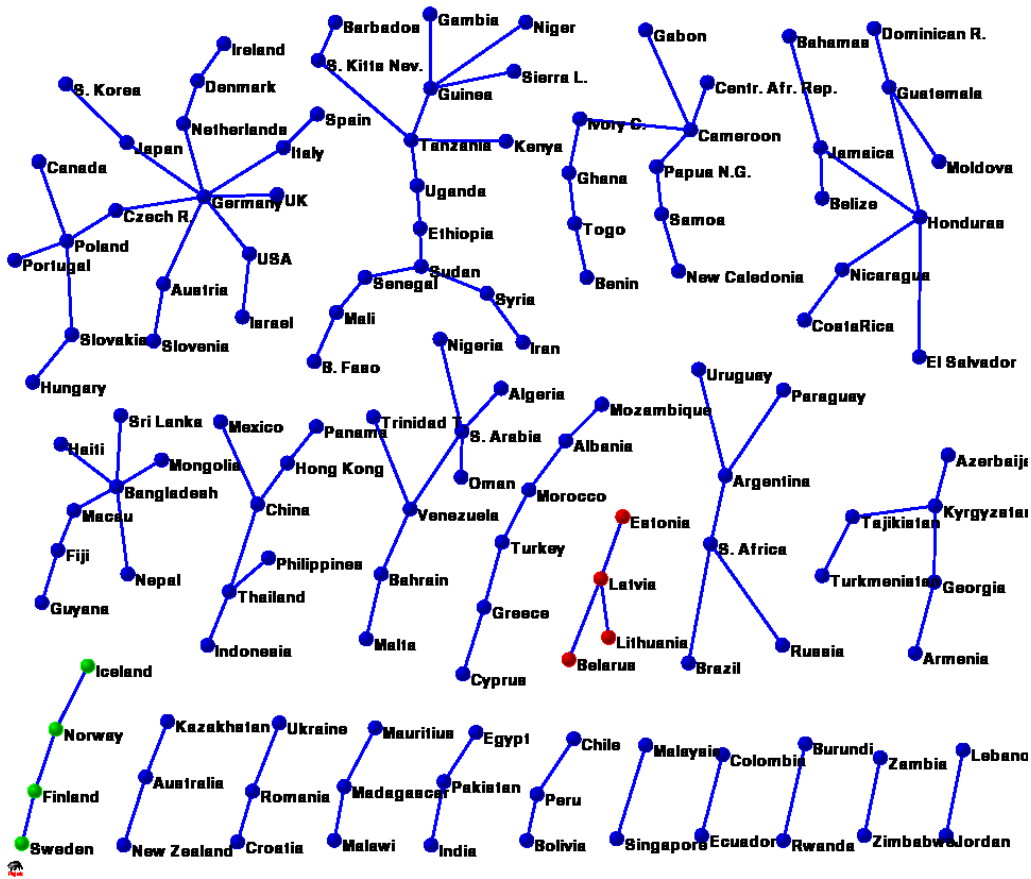}
\caption{\textbf{The Minimal Spanning Forest for the Countries.} The
  various subgraphs have a distinct geographical similarity. We show
  in green northern European countries and in red the ``Baltic''
  republics.  In general neighboring (also in a social and cultural
  sense) countries compete for the production of similar goods.}
\label{Fig2}
\end{figure}

\begin{figure}
\includegraphics[scale=.4]{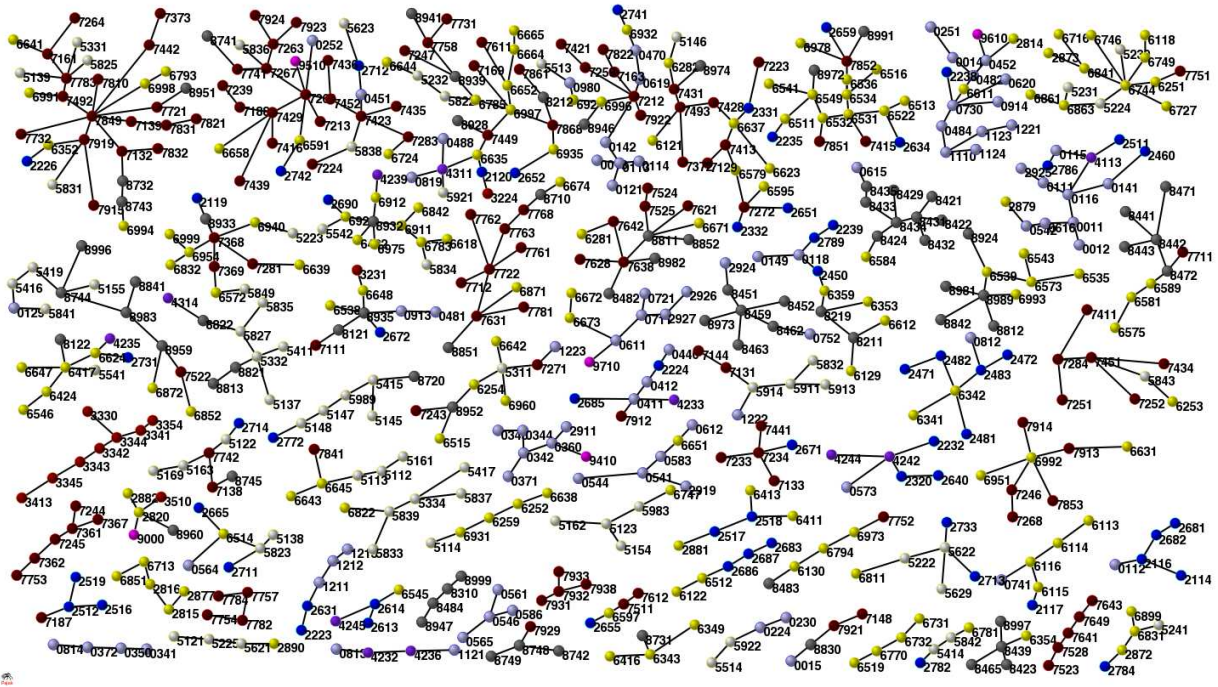}
\caption{\textbf{The Minimal Spanning Forest (MSF) for the Products.}
We put a different color according to the first digit used in COMTRADE classification. 
This analysis should reveal correlation between different but similar products. }
\label{Fig3}
\end{figure}

\begin{figure}
\includegraphics[scale=.5]{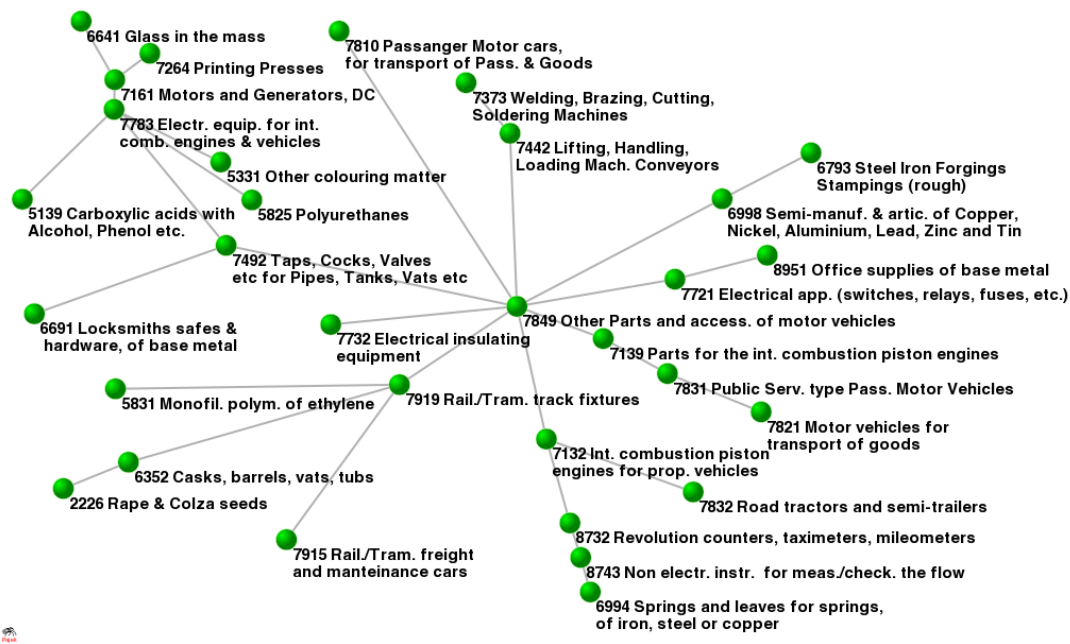}
\caption{\textbf{The largest tree in the Products MSF.} When passing from classification colors 
to the real products name, we see they are all strongly related. It is interesting the presence of colza seeds in the 
lower left corner of the figure.}
\label{Fig4}
\end{figure}


\begin{figure}
\includegraphics[scale=1.0]{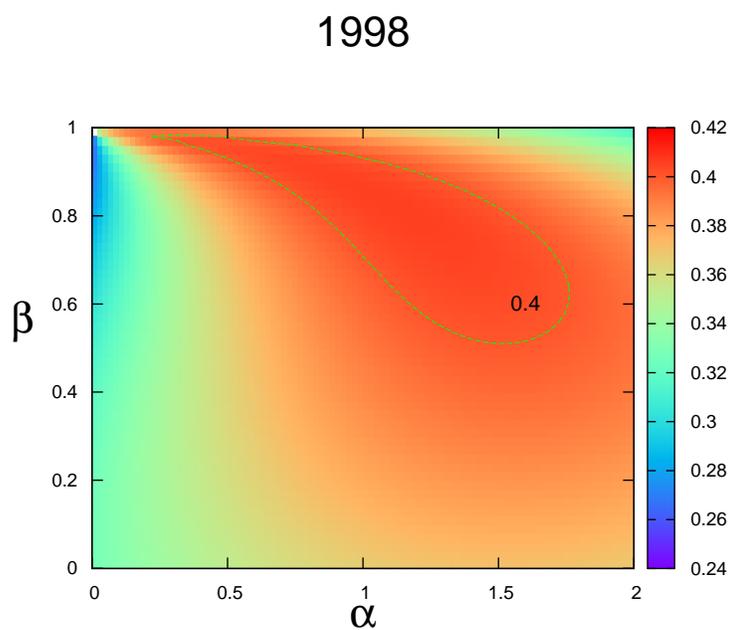}
\caption{\textbf{The plot of the mean Correlation (square of Pearson coefficient, $R2$) 
between logarithm of GDP and
 fixpoint weights of countries in the biased (Markovian) random walk method as a function of
 parameters  $\alpha$ and $\beta$.} The contour plot for a level of $R2=0.4$ is indicated 
as a green loop in the orange region.}
\label{Fig7}
\end{figure}



\end{document}